| Title | **Targeting Cholangiocarcinoma Cells By Cold Piezoelectric Plasmas:** *In Vitro* **Efficacy And Cellular Mechanisms** |
|---|---|
| Authors | Manon Soulier[#*1], Bouchra Lekbaby[#2], Imane Houari[1], Henri Decauchy[1], Allan Pavy[2], Alexia Coumes[1], Romain Morichon[3], Thierry Dufour[†1], and Laura Fouassier[†*2] |
| Affiliations | [1] Laboratoire de Physique des plasmas (LPP), Sorbonne Université, CNRS, Ecole Polytech., Univ. Paris-Sud, Observatoire de Paris, Université Paris-Saclay, PSL Research University, Paris F-75252, France<br>[2] Sorbonne Université, Inserm, Centre de Recherche Saint-Antoine, CRSA, F-75012 Paris, France<br>[3] Cytometry and Imagery platform Saint-Antoine (CISA), Sorbonne Université, Paris F-75012, France<br># co-first author<br>† co-senior authorship |
| Correspondence | thierry.dufour@sorbonne-universite.fr ; laura.fouassier@inserm.fr |
| Ref. | Scientific Reports, Vol. 14, 30178, pp. 16 (2024) |
| DOI | https://doi.org/10.1038/s41598-024-81664-9 |
| Summary | Cold piezoelectric plasma (CPP) is a novel approach in cancer therapy, enabling the development of portable treatment devices capable of triggering cancer cell death. While its effectiveness remains underexplored, this research focuses on its application against cholangiocarcinoma (CCA), an aggressive cancer of the biliary tract. A CPP device is utilized to generate either a corona discharge (Pz-CD) or a dielectric barrier discharge (Pz-DBD) for in vitro experiments. Notably, Pz-CD can deliver more power than Pz-DBD, although both sources produce significant levels of reactive species in plasma and liquid phases. This work shows that CPP causes a gradient increase in medium temperature from the center towards the edges of the culture well, especially for longer treatment times. Although Pz-CD heats more significantly, it cools quickly after plasma extinction. When applied to human CCA cells, CPP shows immediate and long-term effects, more localized for Pz-CD, while more uniform for Pz-DBD. Immediate effects result also in actin cytoskeleton remodeling without alteration of the cell membrane permeability. Long-term effects of CPP, although the antioxidant system is engaged, include activation of the DNA damage response pathway leading to cell death. In conclusion, CPP should be recognized as a promising antitumor therapy. |

# 1. Introduction

Cancer is a leading cause of death worldwide (incidence: 20 million – mortality: 10 million in 2020 [1]). While progress has been made in detecting and treating common cancers like breast and prostate, rare and aggressive cancers like cholangiocarcinoma (CCA) are facing high mortality rates and increasing incidence worldwide [1], [2].

CCA, constituting the second most prevalent malignant liver tumor, comprises intrahepatic (10 to 20%) and extrahepatic (80 to 90%) types [2]. CCA's asymptomatic nature until advanced stages complicates early detection, often leading to biliary tract obstructions or cholangitis [3]. Current curative therapies, mainly surgical removal, are contingent on CCA type, detection stage and patient condition, resulting in only 35% of patients eligible for curative surgery. For the ineligible patients, the first-line treatment includes a palliative chemotherapy (GEMCIS) coupled with an immunotherapy (Durvalumab) [4]. Radiotherapy shows promising results as adjuvant therapy while targeted therapies are employed but rather in second line [5].

Cold atmospheric plasma, a partially ionized gas under ambient conditions, may constitute a breakthrough technology for cancer treatment. Plasma exhibits both high rates of reactive oxygen species (ROS), reactive nitrogen species (RNS) and physical properties (electric, radiative, and thermal) that can interact with biological substrates [6], [7]. Since the 2000s, various configurations of homemade and commercial plasma sources have been tested in a wide number of cancer preclinical models, from cell-based to murine models [8]. Dielectric barrier discharges (DBD) generated by alternative currents have been explored, ranging from plan to jet configurations [9]. Plan-DBD, requiring accessibility to the tumor site, have demonstrated significant antitumor effects [10] showing cytotoxicity specifically on cancer cell lines, compared to their healthy counterpart, depending on the cellular ROS content [11]. Plasma jets involve the propagation of plasma over several centimeters, with support from a neutral gas [12]. Experiments conducted on various tumor models have shown promising results in inducing cell death [13] and tumor growth reduction [14], with mechanistic studies underway to understand the signaling pathways leading to cell death or dysfunction [15-18]. While showing slightly less tumor regression than chemotherapy, plasma treatments may offer advantages by avoiding systemic side effects [19]. Combined therapies involving plasma are also being explored to optimize treatment efficacy and to reduce the required drug doses for chemotherapy [20-21].

However, challenges persist in deploying plasma treatments in hospitals. Direct treatments are currently limited to melanomas or post-surgery interventions and studies are ongoing to develop endoscopic plasma sources to broaden application possibilities [22-24]. Space requirements also affect plasma generators, leading researchers to investigate alternative technologies, such as piezoelectric devices. Existing plasma sources, such as DBDs and jets, which have demonstrated efficacy in inducing cancer cell death, are powered by low-frequency or kilohertz generators that produce stochastic sinusoidal or pulsed periodic signals [11, 19]. These conventional generators require numerous electronic components and cooling systems to achieve the necessary high voltages. In contrast, piezoelectric generators can achieve comparable electrical properties within a compact, portable device, utilizing piezoelectric materials to generate kilovolts efficiently. Resonant piezoelectric transformers consist of the combination of a reverse and direct piezoelectric effect on a piezoelectric crystal, leading to plasma generation at its output [25]. Cold Piezoelectric Plasmas (CPP) are studied extensively, both for their fundamental physicochemical properties at atmospheric pressure [26, 27] and for their practical applications in material science and biology [28-34]. Additionally, these plasmas generate high levels of ROS and RNS able to induce cell death for cancer therapy, but so far their effects have poorly been explored in the treatment of cancer [35-38].

This contribution investigates the impact of CPP on cancer cell viability, considering CCA cells as an in vitro experimental model of cancer. CCA is a highly chemical-based treatment-refractory cancer. To date, CPP has not been considered for the treatment of CCA. In this study, two CPP sources are utilized to generate either a dielectric barrier discharge (Pz-DBD) or a corona discharge (Pz-CD). This work initially characterizes the plasma-liquid interaction between CPP and CCA cells' culture media, based on electric, chemical and thermal properties. The main differences among the two plasmas are relative to thermal dissipation inside the liquid. Subsequently, investigations are being carried out on two CCA cell lines to understand how CPP affects the cell layer, including its spatial distribution and cell viability. The cell layers' observation indicates whether the effects of CPP are immediate or long-term, since both induce a reduction in cell viability 24 h after treatment. Finally, the implications of CPP on cell membrane permeability and actin cytoskeleton, and signaling of DNA damage and repair are further examined, providing insights into the potential applications of CPP in CCA therapy [39].

# 2. Results

## 2.1. Electrical and chemical properties of cold piezoelectric plasma

First, an electrical characterization of the CPP (**Fig. 1a**), either in Pz-DBD (**Fig. 1b**) or Pz-CD (**Fig. 1c**) configuration is realized, followed by an analysis of the reactive species produced in both plasma and liquid phases.

As far as the distribution of electric field lines from the source cannot be investigated in medium or cells, CPP voltage and current signals are measured on a target mimicking human body impedance ($R_T$=1.5 k$\Omega$ – $C_T$=100 pF [40]). In **Fig. 1d**, Pz-DBD provides various heterogeneous periodic micro-discharges characterized by a maximum voltage of 15 V associated with a 240 ns-FWHM (full width high maximum), and a maximum current intensity of 55 mA coupled with a 30 ns-FWHM. As a result, Pz-DBD deposits a relatively low mean power on the target (13 ± 5 mW). In the other configuration, Pz-CD generates intense filamentary discharges. The maximum voltage is 3 times higher than the previous one (51 V) and its FWHM is 2.5 times larger (590 ns). The same order of magnitude is recorded for maximum current intensity (57 mA) but its FWHM is 8 times larger (240 ns). Pz-CD delivers to the target a mean power as high as 74 ± 15 mW.

As expected for a mixed helium-air plasma at atmospheric pressure, nitrogen elements are mostly detected in the spectra of the two plasmas (**Fig. 1e**). First and second orders of nitrogen second positive system excited state ($N_2^* - 1^{st}$ order: maximum at 337.2 nm and 2nd order at 674.4 nm) are predominant. The maximum emission of the first negative system of the nitronium ion ($N_2^{+*}$) is detected in Pz-DBD spectrum only at its maximum (391.5 nm), but not observed for Pz-CD. Nitrite ($NO_2^-$) and hydrogen peroxide species ($H_2O_2$) produced in culture media exposed to CPP are quantified by colorimetric titration (**Fig. 1f**). A relevant increase of nitrites content is observed for the two plasmas, leading to a threshold after 3 min of CPP exposure. The threshold reaches 660 ± 44 μmol/L for Pz-DBD and 915 ± 68 μmol/L for Pz-CD. Concerning hydrogen peroxide, the increase is a little bit slower, with concentrations of 262 ± 86 μmol/L for Pz-DBD and 242 ± 63 μmol/L for Pz-CD after a 5-min treatment. There is no significant difference in $H_2O_2$ concentration determined for the two sources since 1 min of CPP treatment.

The electric characterization reveals that Pz-CD provides a higher voltage to the target for a same maximum current intensity than Pz-DBD. Moreover, the FWHM shows a longer exposure of Pz-CD plasma leading to a five times higher deposited power, increasing potential electric risks. The chemical study demonstrates the production of nitrogen species in plasma phase, able to interact with the chemical compounds of culture media. Compared to untreated medium, nitrite and hydrogen peroxide species are increased whatever CPP condition. Furthermore, nitrites reach higher concentration thresholds with increasing treatment time when medium is exposed to Pz-CD, while hydrogen peroxide is in similar proportions for the two sources.

## 2.2. Cold piezoelectric plasma induces localized thermal effects

**Fig. 2a** and **2b** illustrate the spatial temperature distribution along a radius of the well, from the center to the edge, recorded at 0, 0.5, 1, 2, 3 and 5 min upon CPP treatment. Regardless of the source or duration of exposure, temperatures are higher beneath the nozzle, mostly located at the center of the well, and decrease toward the edges. Considering 1 min of CPP treatment as an example (**Fig. 2b**, blue lines), the temperature at a distance of 10 mm from the center (**Fig. 2b**, dot lines) decreases by 4 °C for Pz-DBD (from 28 °C to 24 °C) and as high as 8 °C for Pz-CD (from 36 °C to 28 °C).

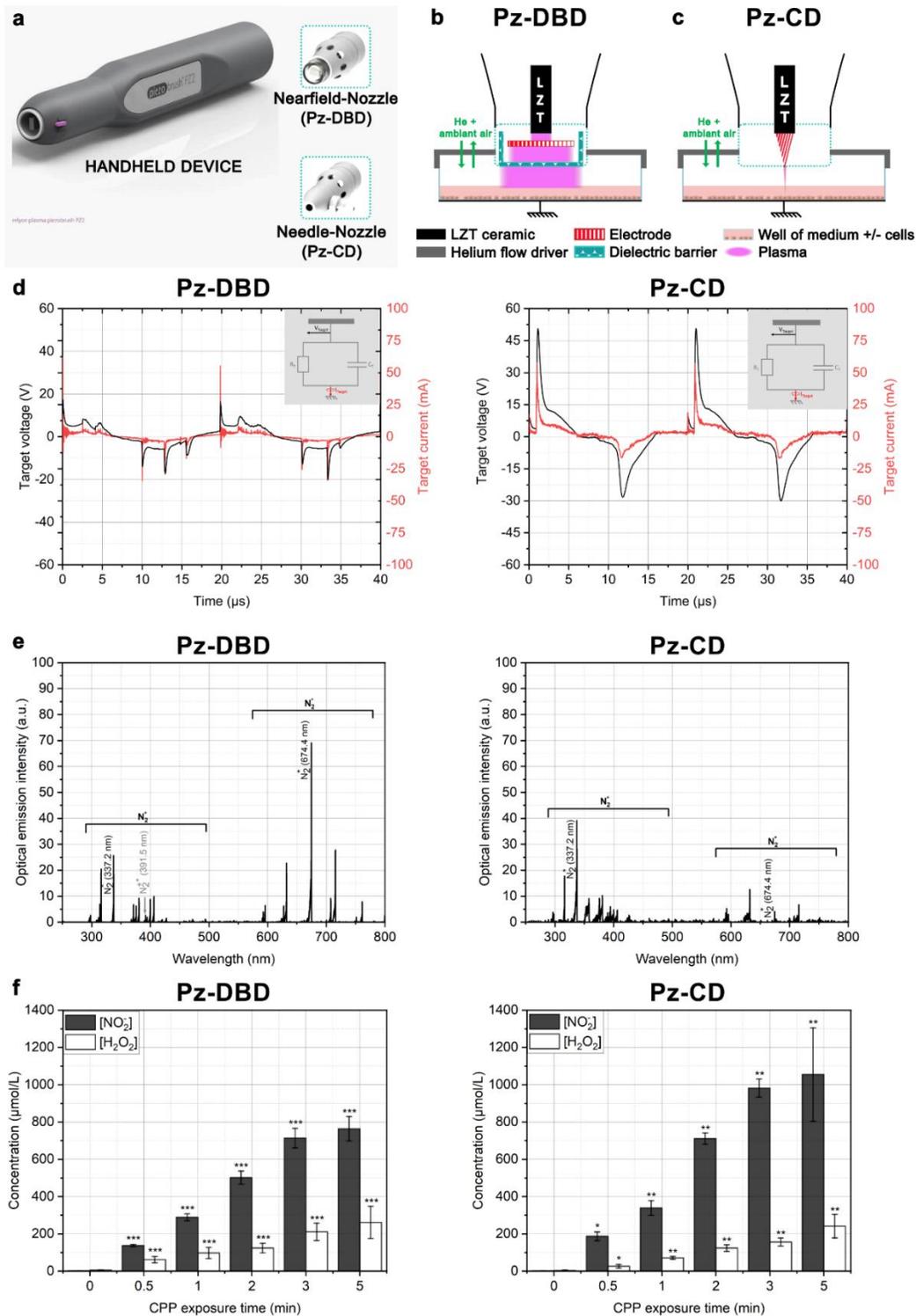

**Fig. 1.** Cold piezoelectric plasma (CPP) device, electric and chemical properties. (a) Photograph of the commercial source of plasma, and the nozzles used in this study. Schemes representing the setups during a plasma treatment of medium with or without cells: (b) for Pz-DBD, the nearfield-nozzle is added; and (c) for Pz-CD, it is the needle-nozzle. LZT ceramic is the Lead Zirconate Titanate piezoelectric material. In both cases, a helium flow driver is inserted to fill the well and maintain the device. (d) Instantaneous voltage and current intensity recorded on a target mimicking human body impedance ($R_T = 1.5$ $k\Omega - C_T = 100$ pF [40]). The target and probe location are schematized in insert. (e) Optical emission spectra from CPP sources interacting with culture medium in a well. Nitrogen species are in majority, mainly represented by the second positive system (respectively, first and second orders) of molecular nitrogen $N_2^*$ and the first negative system of the nitronium ion $N_2^{+*}$. (f) Nitrite and hydrogen peroxide content in culture media after CPP exposure from 0 to 5 min. Values are expressed as means ± s.e.m. from at least three replicates. Significant differences for *, $p<0.05$; **, $p<0.01$; ***, $p<0.001$ are compared with untreated cells (0 min).

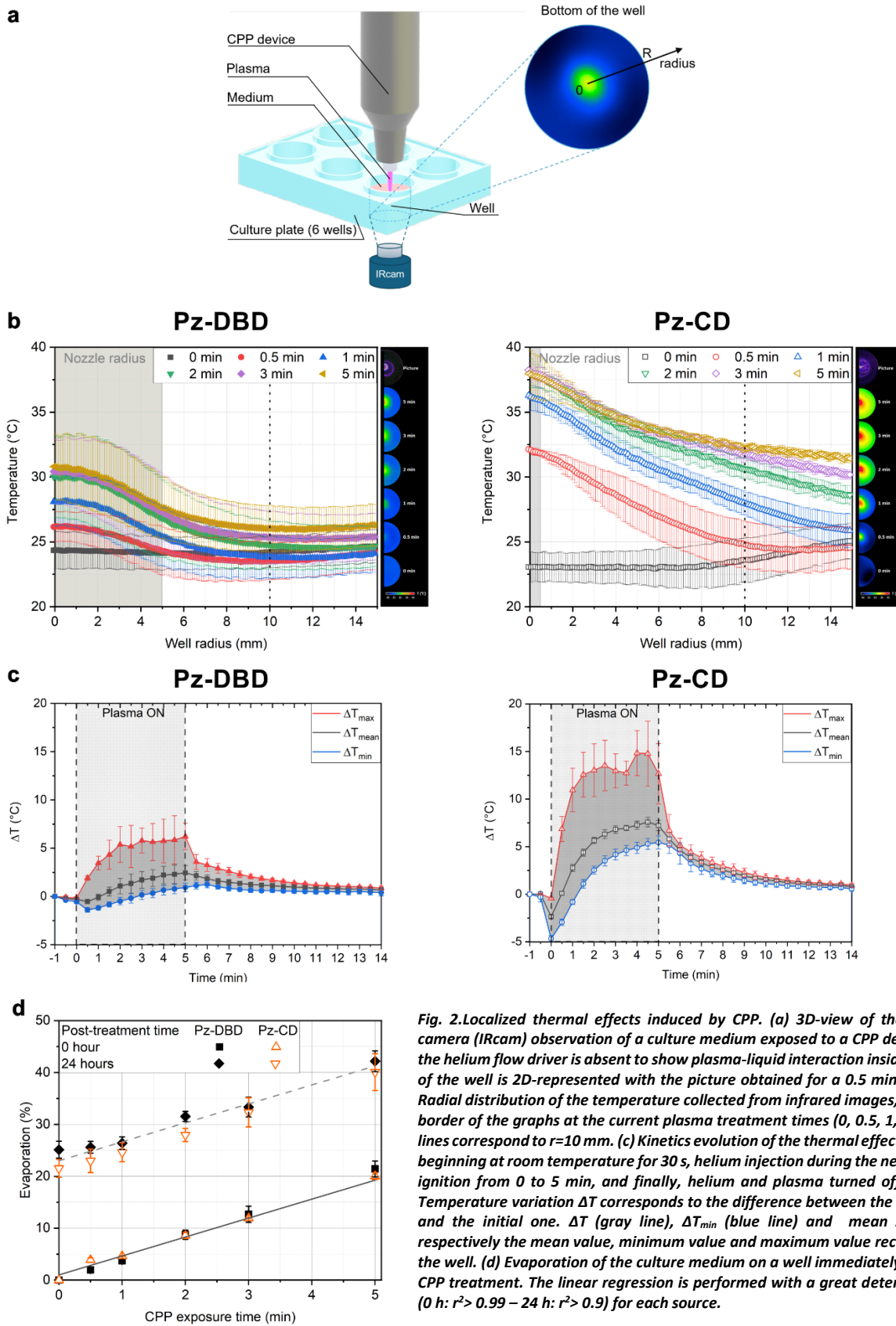

**Fig. 2.** Localized thermal effects induced by CPP. (a) 3D-view of the setup for infrared camera (IRcam) observation of a culture medium exposed to a CPP device. In this scheme, the helium flow driver is absent to show plasma-liquid interaction inside a well. The bottom of the well is 2D-represented with the picture obtained for a 0.5 min-Pz-CD exposure. (b) Radial distribution of the temperature collected from infrared images, aligned on the right border of the graphs at the current plasma treatment times (0, 0.5, 1, 2, 3 and 5 min). Dot lines correspond to r=10 mm. (c) Kinetics evolution of the thermal effect for the two sources, beginning at room temperature for 30 s, helium injection during the next 30 s; then, plasma ignition from 0 to 5 min, and finally, helium and plasma turned off for the last 9 min. Temperature variation ΔT corresponds to the difference between the current temperature and the initial one. ΔT (gray line), ΔT$_{min}$ (blue line) and mean ΔT$_{max}$ (red line) are respectively the mean value, minimum value and maximum value recorded on the area of the well. (d) Evaporation of the culture medium on a well immediately (0 h) and 24 h after CPP treatment. The linear regression is performed with a great determination coefficient (0 h: r$^2$> 0.99 − 24 h: r$^2$> 0.9) for each source.

**Fig. 2c** depicts the temperature kinetics of the two sources, tracking the heating and cooling phases. Before CPP exposure, the temperature decays due to helium injection, with a more pronounced decrease observed for Pz-CD (drop of −5 °C) than for Pz-DBD (drop of −1.5 °C). During the heating phase ("plasma ON"), the temperature increases quickly to a higher threshold for Pz-CD (from 2 min: +15 °C for Pz-CD *vs*. + 5 °C for Pz-DBD). The post-CPP cooling phase shows temperatures converging rapidly to room temperature ($\Delta T = 0$ °C), even without helium flow. Specifically, Pz-DBD shows a 3 °C-decrease within just 1 min after plasma extinction, while Pz-CD experiences a quick 7 °C drop.

Furthermore, the evaporation of culture medium after each CPP treatment (0, 0.5, 1, 2, 3 and 5 min), is assessed immediately (0 h) and after 24 h (**Fig. 2d**). Initially, evaporation appears to be higher for Pz-CD treatment, recording a weight percentage (%$_{wt}$) of 4%$_{wt}$ against 2%$_{wt}$ for Pz-DBD at 0.5 min. However, for longer exposure durations, both sources converge to a similar evaporation rate of 20%$_{wt}$ after 5 min. A global linear regression can be applied with a strong determination coefficient ($r^2 > 0.99$) for the two sources. Evaporation reaches 40%$_{wt}$ after 24 h, maintaining the trend observed immediately post-treatment ($r^2 > 0.9$).

This thermal characterization shows an overall increase in the temperature of the medium provided by the CPP sources, mainly located below the nozzles and diffusing towards the well edges while increasing treatment time. Notably, Pz-CD induces a more significant increase in temperature compared to Pz-DBD. Kinetics experiments also demonstrate a quick cooling of the medium after plasma extinction. This investigation also exhibits a linear variation of evaporation with exposure time, regardless of the CPP source. This phenomenon is persistent in time, following the same regression slope immediately and 24 h after CPP treatment, highlighting the control and efficiency of the process across both plasma sources.

## 2.3. Effects of cold piezoelectric plasma on cancer cells are spatiotemporally dependent

In the next experiments, cholangiocarcinoma cell lines, HuCCT-1 and EGI-1, are treated with CPP. To differentiate the immediate vs. long-term effects triggered by CPP on cells within the well, analyses are conducted at two distinct timescales, immediately (0 h) or 24 h after CPP treatment. These effects are evaluated on the cell layer by counting the cells identified with a blue-fluorescent DNA stain (**Fig. 3a**).

The first case, called 0 h in **Fig. 3b**, corresponds to the observation of the cell layer immediately after CPP treatment from 0 to 5 min. No modification of HuCCT-1 cell layer distribution is observed after Pz-DBD treatment, regardless of the exposure time and the number of cells is unchanged. The same observations apply to EGI-1 cells exposed to Pz-DBD. On the contrary, the layer of HuCCT-1 cells treated by Pz-CD for 3 and 5 min displays a high reduction of the cell number below the plasma impact region and it is expanding with treatment time. Such layer modification is not observed for EGI-1 cells exposed to Pz-CD.

The second case, called 24 h in **Fig. 3c**, corresponds to the observation of the cell layer 24 h after CPP treatment from 0 to 5 min. From 0 to 2 min of Pz-DBD treatment, HuCCT-1 cell layer is quasi-homogeneous and a decrease in cell number is observed. After a 2 min-Pz-DBD treatment, only 30% of cells are still observed on a radius of the layer. The cell number decreases mainly bellow the DBD nozzle, leaving less than 10% cells at the 3 and 5 min treatment times. The same observations can be done for EGI-1 cells exposed to Pz-DBD with a slight decrease in cell number (at 2 min: still 50% of cells on a radius). Pz-CD treatments provide a decrease of HuCCT-1 cell number around the plasma impact as early as 0.5 min of exposure, and this reduction expands until 3 min of treatment. After 5 min of exposure, the layer is quasi non-existent with only 0.3% of cells observed on a radius of the layer. Similar to Pz-DBD treatments, EGI-1 cells exhibit a slight response 24 h following Pz-CD exposure.

This experiment highlights an immediate effect of Pz-CD on HuCCT-1 cell layer, becoming noticeable after 3 min of treatment, but not on EGI-1 cells. Conversely, Pz-DBD treatment does not exhibit any immediate effect on each cell line. In contrast, long-term effects are visible; the two sources lead to a decrease in cell number 24 h after CPP treatment, indicating an impact of CPP on cell viability.

## 2.4. Cold piezoelectric plasma reduces cell viability by inducing DNA damage response-linked apoptosis

The viability of the two human CCA cell lines is assessed 24 h after their exposure to each CPP source (**Fig. 4a**). The two sources elicit similar responses, showing survival regressions for the two cell lines. If Pz-DBD treatments result in a significant decrease in cell viability, HuCCT-1 cells exhibit higher sensitivity to CPP, with over 75% cell death after a 5 min-treatment, compared to approximately 55% for EGI-1 cells. As expected, IC50 values are reached before 5 min, specifically between 2 and 3 min for HuCCT-1 cells and between 3 and 5 min for EGI-1 cells. On the other hand, Pz-CD treatments also demonstrate a notable reduction in cell viability, growing with exposure time. HuCCT-1 cells respond similarly to the two sources while EGI-1 cells are more adversely affected by Pz-CD, with an IC50 value reached at 3 min. Therefore, from 3 min of treatment, Pz-CD seems more efficient to reduce EGI-1 cell viability. These data indicate that CPP, regardless of the type of discharge, decreases the viability of CCA cells, even EGI-1 cells which are more resistant than HuCCT-1 cells. Based on this viability study, treatment times 2–3 min for HuCCT-1 and 3–5 min for EGI-1 are selected to decipher DNA damage response pathway and antioxidant defenses in cells.

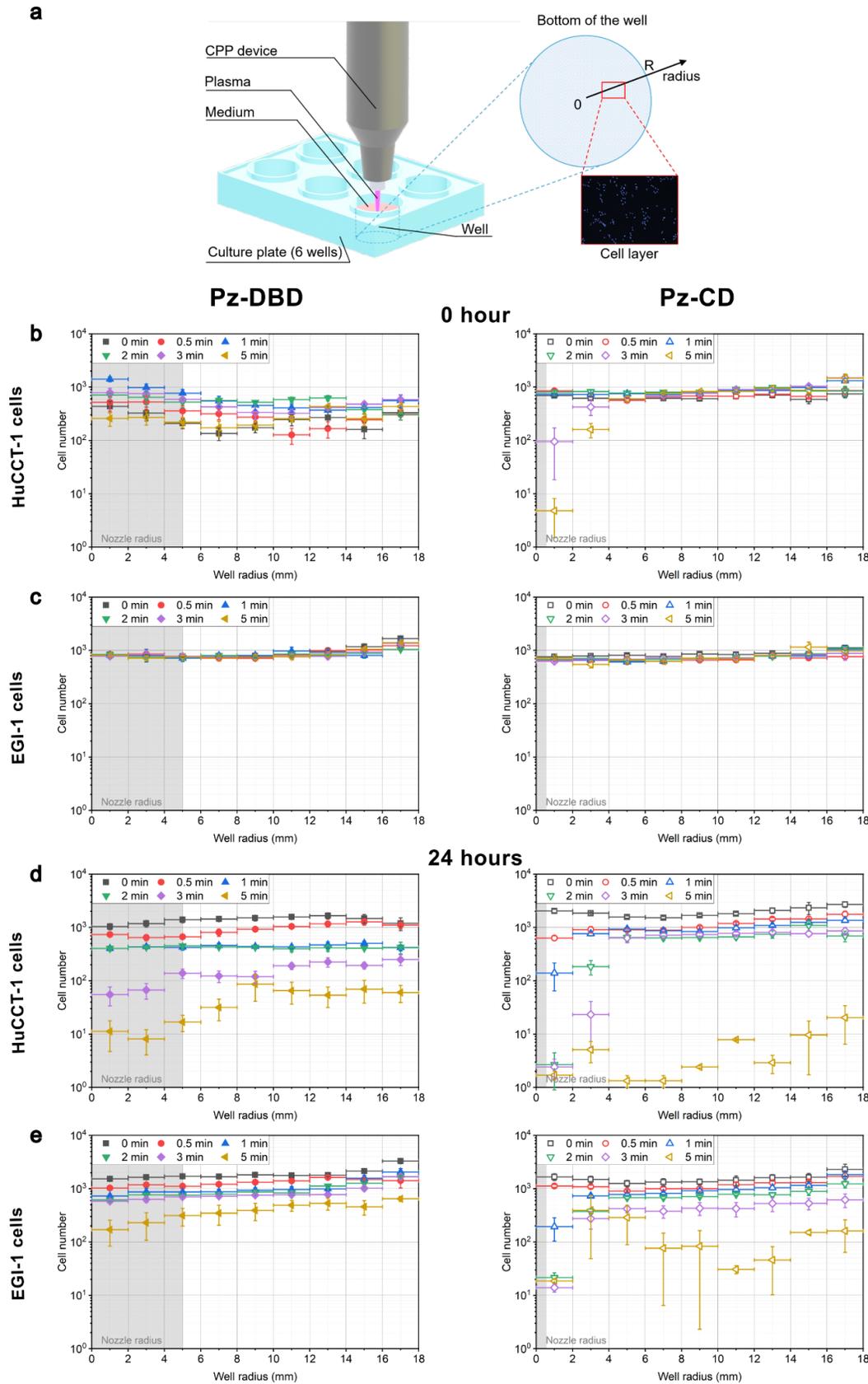

**Fig. 3. Spatiotemporally dependence of CPP effects on the CCA cell layer distribution.** (a) 3D-view of the setup when the cell layer is exposed to a CPP device. In this scheme, the helium flow driver is absent to show plasma-liquid interaction inside a well. The bottom of the well is 2D-represented to focus on the cell nucleus observed from DAPI fluorescence images acquired along a radius of the well. Cell distribution observed on a radius of the well after CPP exposure, immediately (0 h) either for (b) HuCCT-1 cells or (c) EGI-1 cells. The same conditions are then observed 24 h after treatment and quantified for the two cell lines: (d) HuCCT-1 and (e) EGI-1. The nozzle radius is notified by gray background, respectively 5 mm for Pz-DBD and 0.5 mm for Pz-CD. Cell number is plotted in log-scale.

In the cells, CPP exposure leads to DNA breaks followed by a DNA damage response (DDR) coordinated by several key proteins. Three of them are studied including γH2AX, the phosphorylated form of histone H2AX, the checkpoint kinase 1 (Chk1) and the tumor suppressor p53. One of the earliest cellular responses to DNA damage is driven by γH2AX that serves as a checkpoint for DNA repair. While γH2AX is not observed by immunofluorescence in untreated cells, foci of γH2AX are widely visible in almost all cells treated either with Pz-DBD (**Fig. 4b**) or with Pz-CD (***Supp. Fig.a***), regardless of the cell line. The effect of CPP on γH2AX is independent of the plasma source; no difference in γH2AX immunostaining is observed in CPP-treated cells between Pz-DBD and Pz-CD. Expression of γH2AX is confirmed by western blot in CCA cells treated with

Pz-DBD (**Fig. 4c**) or Pz-CD (***Supp. Fig.b***). Chk1 is a major kinase phosphorylated in response to DNA damage. Upon CPP treatment, phosphorylation level of Chk1 increases in CCA cells. One target of the kinase Chk1 is p53. As shown in **Fig. 4c**, p53 is phosphorylated in CCA cells subjected to CPP treatment, triggering cell death, as evidenced by the expression of cleaved caspase 3 and cleaved PARP (**Fig. 4c** and ***Supp. Fig.b***).

The following assessments focus on the regulation of the antioxidant enzymes by CPP. The key antioxidant enzymes including MSRB3, SOD1, SOD2, CAT2 and HMOX1 (HO-1) are analyzed at mRNA levels by RT-QPCR. The expression of these enzymes is increased in cells treated with Pz-DBD, with the exception of *CAT2* that is decreased (**Fig. 4d**). In contrast, Pz-CD has less impact on antioxidant defense than Pz-DBD; it only induces an increase in *HMOX1* expression, with no change in the others (***Supp. Fig.c***).

In conclusion, CPP reduces cell viability by activating a DNA damage response pathway leading to cell death, although the antioxidant system is mostly engaged for Pz-DBD and not for Pz-CD.

## 2.5. Cold piezoelectric plasma induces cytoskeleton remodeling without affecting cell membrane permeability

CPP's electric field could be responsible for cell permeabilization which, in turn, could mediate cell viability reduction [41]. To verify this hypothesis, the permeability of the cell membrane is assessed using a fluorescent cell-impermeant dye. As seen in **Fig. 5**, only the immediate effect (0 h) is taken into account, to focus on the plasma electric effects. The cells pre-incubated with the fluorescent cell-impermeant compound are immediately observed after CPP treatments during 2 and 3 min for HuCCT-1 and 3 and 5 min for EGI-1 cells. As shown in **Fig. 5a**, no dye uptake is observed in cells treated with Pz-DBD or with Pz-CD (data not shown), compared with the positive control which consisted of treating the cells with Triton, a detergent known to induce pores in the membrane.

In light of this result, the effects of CPP on the intracellular environment are assessed, in particular on the actin cytoskeleton that plays a major role in the maintenance of epithelial barrier and morphology [42]. CPP is applied to cells under the same conditions as for the permeability study, followed by immediate analysis of the actin cytoskeleton structure (**Fig. 5b**). Depending on the CCA cell line, actin cytoskeleton rearrangements differ in response to plasmas. In untreated HuCCT-1 cells, F-actin outlines the cell membrane and a few actin fibers are visible in the cytoplasm. Plasma treatment, either Pz-DBD or Pz-CD, for 2 min induces a drastic remodeling of actin cytoskeleton with the formation of prominent intracellular actin stress fibers and of membrane protrusions. After 3 min of exposure, F-actin staining becomes more intense at cell membrane level, with the formation of more intense and numerous protrusions. Regarding EGI-1, untreated cells are cohesive and display a staining of F-actin at the plasma membrane without fibers visible into the cytoplasm. Upon a 3-min Pz-DBD treatment, cell shape is changed with a wider cytoplasm and a loss of F-actin at the cell-cell junctions. After 5 min of Pz-DBD exposure, cells become deformed with a more elongated cytoplasm. In contrast, Pz-CD induces a different actin behavior, with a reinforcement of filamentous actin at the cell membrane, without formation of membrane protrusion or stress fibers.

Overall, these data indicate that both plasmas have short-term effects on cell phenotype by reorganizing the actin cytoskeleton, which differs between CCA cell lines and have virtually no impact on cell permeability.

## 3. Discussion

The present work investigates the effects of two cold piezoelectric plasma (CPP) sources in the purpose of cancer therapy, using human cholangiocarcinoma (CCA) cells as in vitro models. CCA is a rare cancer of the biliary tract that displays high mortality rates and increasing resection incidence, without other curative treatment than tumor resection for eligible patients [2]. As reviewed by Korzec et al. [25], CPP sources are currently used for material surface treatments, exposing directly the piezoelectric ceramic to a dielectric surface (ceramic or polymer) in ambient air for surface activation, increasing wettability or adhesion before printing, painting or coating processes. They highlighted ozone production by direct CPP, useful for disinfection of liquid, material and biological substrates. More recently, Konchekov et al. [43] as well as Korzec et al. [25] (the latter designing the commercial device piezobrush® PZ2 (**Fig. 1a**)) have implemented nozzles to perform corona or dielectric barrier discharges for conductive, liquids or biological substrates, using CPP as a plasma bridge.

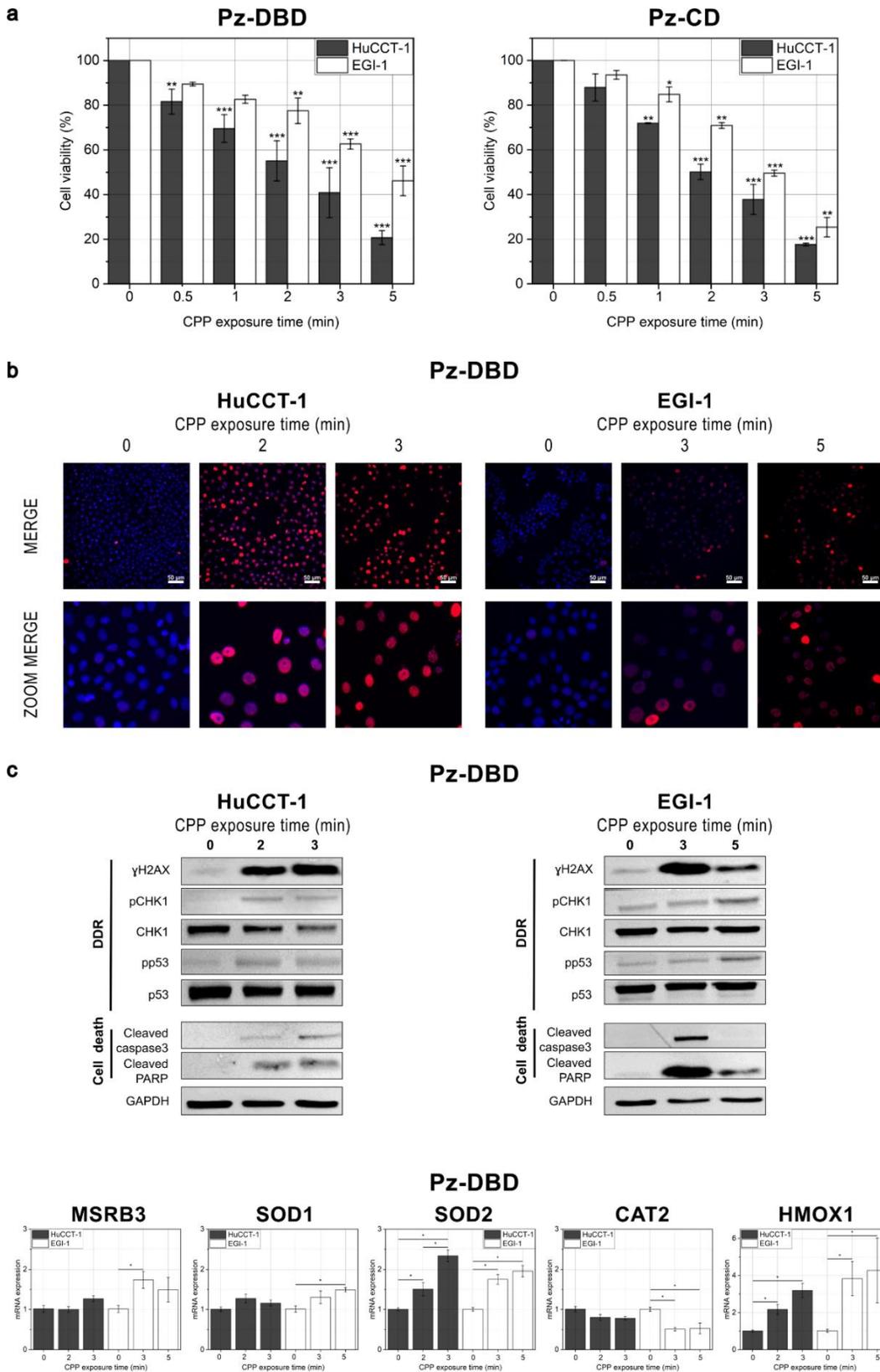

Fig. 4.Effects of CPP on cell viability, DNA damage response, cell death and antioxidant system. (a) HuCCT-1 and EGI-1 CCA cells viability is quantified 24 h after CPP exposure for 0.5, 1, 2, 3 and 5 min by crystal violet assay. (b-c) CCA cells are treated with Pz-DBD for 2 and 3 min for HuCCT-1 – 3 and 5 min for EGI-1 and analyses are performed 24 h post-treatment. (b) Representative images of γH2AX analyzed by immunofluorescence in cells. (c) Representative images of western blot analysis of γH2AX, phosphorylated and total Chk1, phosphorylated and total p53, cleaved caspase-3 and cleaved PARP. The samples derive from the same experiment and gels/blots are processed in parallel. Blots are cropped; original uncropped blots are presented in Supplementary materials. (d) Expression of MSRB3, SOD1, SOD2, CAT2 andHMOX1 at mRNA level by RT-QPCR. Values are expressed as means±s.e.m. from at least three independent cultures. Significant differences for *, p<0.05; **, p<0.01; ***, p<0.001 are compared with untreated cells (0 min).

A few studies, conducted by two research groups [35-38], investigate the same CPP source for cancer treatment but in diverse discharge configurations and cancer types. Suzuki-Karasaki et al. employ the piezoelectric ceramic without nozzle to generate an "Air Plasma Activated Media" (APAM) on osteosarcoma and oral cancer [37-38], favoring ozone production to trigger oxidative cell death caused by hydrogen peroxide and iron. As well as they do not expose directly the cells in culture medium, they use mostly long-lifetime ROS produced by plasma-medium interaction [44]. The APAM exhibits the tumor-selective anticancer activity, by increasing the mitochondrial ROS rates in cells. On the other team, Wang et al. are using the device with a needle-nozzle, aiming to generate corona discharges in ambient air for the treatment of colon and lung cancer [35] or hepatocellular carcinoma [36]. Based on the imbalance of high oxidative stress and low antioxidant capacity involved in cancer cells, they highlight the effects and mechanisms (apoptosis and autophagy) of ROS, assuming that they do not investigate any electromagnetic, radiative or thermal effect.

The present work corresponds to the first study deciphering the immediate and long-term effects caused by two CPP in a mixed helium-air atmosphere on CCA cells. The sources, designed to generate either a dielectric barrier discharge (Pz-DBD, **Fig. 1b**) or a corona discharge (Pz-CD, **Fig. 1c**), are incremented with helium flow drivers for in vitro exposure of CCA cells in 6-well plates. Helium is used as a working gas to ignite uniform discharges in larger gaps without increasing gas temperature thanks to its inert chemical nature [45]. A benchmark of the sources establishes that they exhibit similar effects on CCA cells for the first two minutes of exposure. Then, Pz-CD becomes a more efficient but also a more hazardous in vitro process, while Pz-DBD is still efficient and safe. These phenomena are mostly explained by the discharge physicochemical properties. A characterization of the plasma-liquid interaction is associated with cellular, molecular and DNA responses to decipher the immediate and long-term effects of CPP exposure.

Since CPP is generated by a commercial device, its electrical characterization can only be achieved indirectly, through the exposure of a conductive target, mimicking the human body impedance [40]. Even if the mean power deposited by Pz-CD is five times higher than the one deposited by Pz-DBD, none of them show irreversible electric risk for cell treatment [46]. Indeed, during a 20 μs-period of the piezoelectric direct effect (**Fig. 1d**), Pz-DBD provides at least three positive micro-discharges from the dielectric barrier to the target with a decreasing maximum intensity of a few tens of ns, while Pz-CD generates a single filamental positive discharge of a few hundreds of ns directly from the powered electrode to the target. When generating CPP during a few minutes on culture medium, the conductivity is lower than the one of the human body impedance, ensuring no electric irreversible damage. Furthermore, the assessment of membrane permeability on CCA cells (**Fig. 5a**) confirms that there is no penetration of low-molecular weight molecules (630 Da) across the cell membrane, whatever the CPP exposure. The effects of CPP on cell permeability has never been studied in cancer cells. Thereby, this study first evidences that CPP has no effect on membrane permeability analyzed immediately after treatment, suggesting that its electric field has no deleterious effect on membrane integrity and prevents the passage of molecules through the cell membrane. These data contrast with plasma sources other than CPP known to increase cell permeability [47, 48] such as a kINPen MED/argon inducing membrane permeability, assessed by different molecules. These differences in methodology may explain the discrepancies between the studies.

Although no alteration in cell membrane integrity are detected under CPP conditions, notable intracellular alterations are highlighted in the cellular cytoskeleton with remodeling of actin filaments (F-actin) immediately after plasma treatment. These alterations are evidenced by the presence of actin stress fibers, membrane protrusions and changes in the F-actin network at cell-cell junctions. Such description of actin cytoskeleton remodeling in response to cold piezoelectric plasma is novel. The only reports on actin in cells treated with plasma (kINPen MED/argon) provide evidence of a change in globular actin (G-actin) and filamentous actin (F-actin) ratio with an increase of the G-/F-actin ratio [47, 49]. The induction of actin stress fibers has also been highlighted in cancer cells, i.e. Hela cells, in response to plasma jet exposure [50]. These changes in the actin cytoskeleton may reflect the effect of cold plasma electric field since plasma activated media has no effect [50], although ROS/RNS or temperature cannot be ruled out [51]. Actin cytoskeleton is considered as a sensor of cell responses to external cues and its rearrangement may depend on cell phenotype. It is interesting to emphasize that notable differences have been observed between the two CCA cell lines upon CPP treatment. One explanation is that the two cell lines have different phenotypic characteristics, EGI-1 cells displaying a more pronounced epithelial phenotype with stronger expression of epithelial markers compared to HuCCT-1 cells, which express mesenchymal markers [52]. In EGI-1 cells, an enlargement of cell cytoplasm is observed upon Pz-DBD without visible stress fiber formation. In contrast, HuCCT-1 cells show stress fibers and cellular extensions upon Pz-DBD or Pz-CD, in line with their mesenchymal-like features. Overall, irrespective of CCA cell phenotype, cold piezoelectric plasma leads to alterations of the actin cytoskeleton, which is likely to impair cellular functions such as cell migration and adhesion [49, 53].

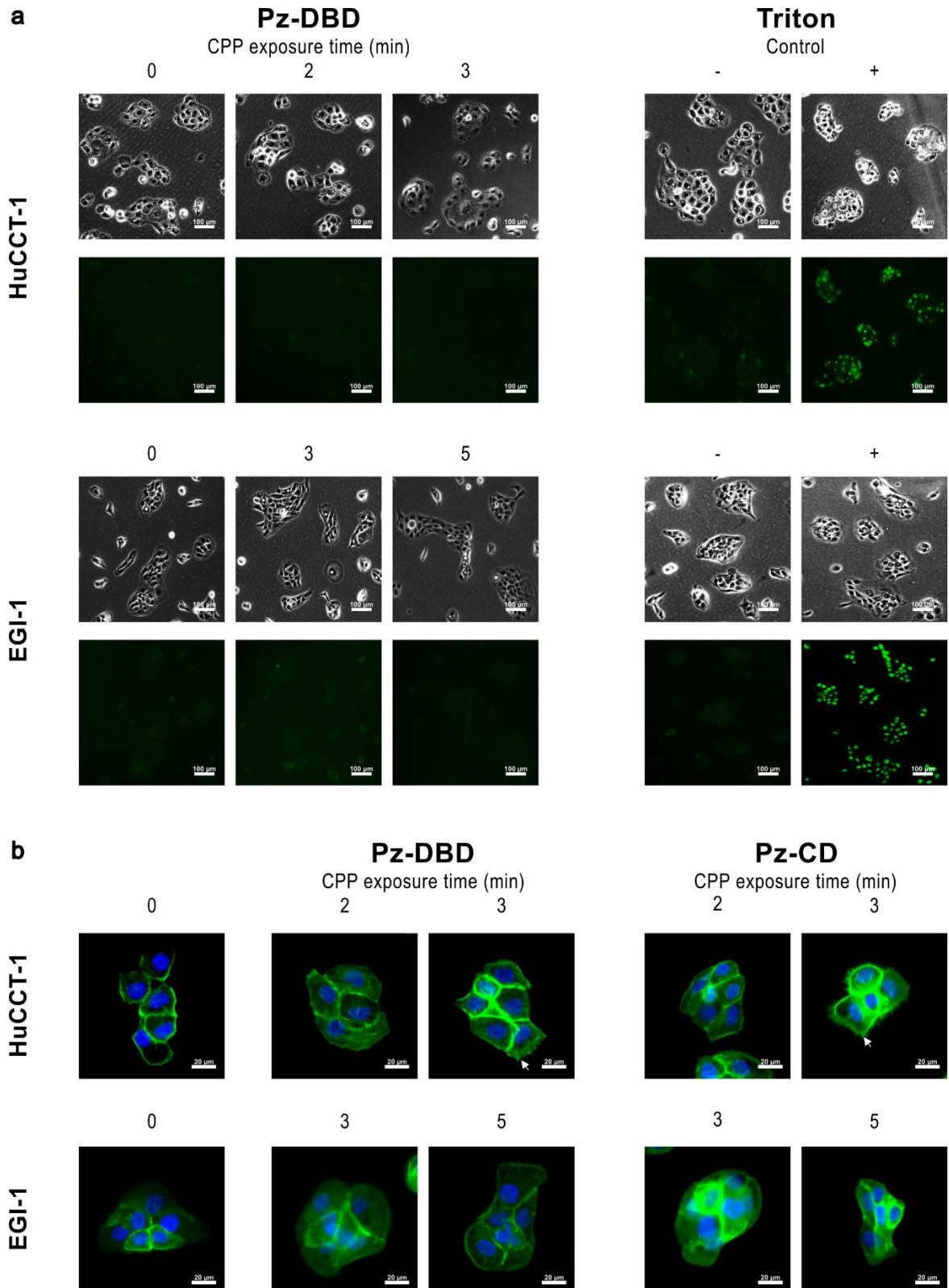

**Fig. 5. Effect of CPP on cell membrane permeability and actin cytoskeleton.** Representative immunofluorescence images of (a) cell impermeant dye and (b) F-actin cytoskeleton in CCA cells captured immediately after Pz-DBD exposure for 0.5, 1, 2, 3 and 5 min. (a) Cell impermeant dye does not cross the cell membrane of CCA cells treated with Pz-DBD compared to cells treated with Triton X-100, a detergent that induces pores in the membrane. (b) F-actin is reorganized in cells treated with plasma. CPP induces actin stress fibers and membrane protrusions in HuCCT-1 but not in EGI-1 cells. In both cells, actin staining is reinforced at the cell membrane upon Pz-CD treatment. White arrows indicate membrane protrusion.

CPP is considered as a cold process in regard to a thermal plasma which is characterized by gas temperature around 10,000 °C. However, CPP can heat higher than room temperature, especially when interacting with a conductive medium. CPP diffuses throughout the medium, facilitating heat transfer until the medium achieves a steady-state thermal equilibrium. In this study, the temperature of culture medium exposed to CPP demonstrates a localized thermal effect below the source, diffusing inside the medium while increasing treatment time. Pz-CD shows a higher rise since the beginning of exposure as much as a quicker diffusion toward the edges than Pz-DBD. None of them reaches temperatures able to induce damages on the cells [54] (**Fig. 2b**, T < 40 °C) in these experimental conditions. Timmermann *et al*. [28] have thermally characterized the same CPP sources to generate pure air plasmas for assessing antimicrobial efficacy. Using the DBD configuration with a 2 mm-gap or the corona discharge up to 7 mm-gap, they reached temperatures higher than 40 °C in less than 30 s on conductive substrates, concluding that the devices cannot be used for medical purposes. Their work confirmed the necessity of using helium-air plasmas when exposing conductive substrates.

However, a special care must be taken when operating with Pz-CD upon treatment longer than 5 min, both for thermal and electric hazards. Even if the maximum temperature of the medium is already reached after 5 min of CPP exposure (and even before, **Fig. 2c**), the evaporation of culture medium increases linearly with exposure time (**Fig. 2d**). Therefore, the conductivity of the substrate increases; the 6-well plate is grounded at the bottom. In the case of Pz-CD, this phenomenon leads to a direct current propagation of a spark discharge to the ground [55] and a rising temperature. There is no similar effect with Pz-DBD. When increasing the evaporation, the gap between the dielectric barrier and the substrate rises, leading to plasma extinction. Evaporation of liquid exposed to cold plasma in in vitro experiments is not well documented but Freund *et al*. [56] utilized the kINPen MED/argon on six different liquids used for clinical purposes and measured also a linear evaporation of 50 mL of liquid exposed during 80 min.

Plasma-liquid interaction cannot be studied without investigating the reactive species in the two phases. Optical emission spectroscopy provides a qualitative information about the excited states present in the plasma phase 1 mm above the culture medium. Excited molecular nitrogen is predominantly detected in the plasma phase, while oxygen and helium emissions are notably absent. This absence can often be attributed to the production of active species which, although generated, remain non-emissive due to quenching. This quenching occurs through mechanisms such as collisions with other molecules, rapid recombination reactions and energy transfers to the surrounding medium (**Fig. 1e**). Indeed, when performing analysis of ROS and RNS in the medium, the rising effects of CPP are significant after 0.5 min of exposure (**Fig. 1f**). Therefore, we cannot exclude the presence of $O_2^-$ or singlet $O_2$ within the discharge, which may still participate in generating reactive species in the liquid phase, potentially contributing to cancer cell cytotoxicity [12, 57, 58]. This work focuses on nitrites and hydrogen peroxide which are long-lived synergetic species involved in cell death mechanisms, mainly known to induce cell apoptosis [37, 59, 60]. No other chemical study is conducted in the same conditions but Konchekov *et al*. [31, 43] reached similar nitrite and hydrogen peroxide species content when exposing directly the piezoelectric ceramic to different liquids (in the order of 100 μmol/L for a few min of exposure) and Timmermann *et al*. [28] obtained concentrations in the range of tens ppm of nitrites produced by the sources in the far field of the air plasma phase.

Both sources show long-term effects, significant after 1 min of CPP exposure for the two cell lines (**Fig. 3d-e**) and correlated with the reduction in cell viability and an increase of cell death (**Fig. 4a** and **4c**). **Fig. 4a** illustrates the effects of Pz-DBD and Pz-CD treatments on cell viability. For each cell line, the differences observed between the two plasma configurations are not statistically significant, with the exception of the EGI-1 cell line at 5 min. At this time point, **Fig. 1f** shows equivalent $H_2O_2$ levels across both configurations, while nitrite concentration differs, measuring 1050 μM for Pz-CD and 750 μM for Pz-DBD. This discrepancy suggests that EGI-1 cells may exhibit a heightened sensitivity compared to HuCCT-1 cells in response to elevated nitrite levels. Overall, these effects depend on the ROS/RNS generated by CPP that induce a cellular oxidative stress affecting DNA integrity as evidenced by DNA strand breaks and DDR response, that are both promoted by the two CPP. These data indicate that common pathways are systematically induced in response of CPP irrespective of the source. However, antioxidant enzyme expression differs from the two sources. While with Pz-DBD, most enzyme expression is up-regulated; unexpectedly, it is not the case for Pz-CD. The explanation for this discrepancy deserves further investigation. It is worth pointing out that Wang *et al*. have nicely shown that CPP decreases antioxidant capacity at a protein level (glutathione) in hepatocellular carcinoma cells, thus aggravating intracellular imbalance [36]. Together, CPP devices, like other plasma sources, have the ability to regulate the antioxidant system.

In conclusion, both CPP sources have antitumor effects on CCA tumor cells by inducing cell death, although they differ in their physicochemical properties. In terms of biological effects, the novelty lies in the remodeling of the actin cytoskeleton, which is very pronounced in tumor cells shortly after CPP treatment, while the integrity of the cell membrane remains unaffected. This study provides evidence that the cold atmospheric plasma delivered by CPP is effective against CCA tumor cells.

# 4. Methods

## 4.1. Cold piezoelectric plasma device and exposure conditions

Plasma device: the piezobrush® PZ2 (**Fig. 1a**) from Relyon Plasma GmbH Company is utilized as a cold piezoelectric plasma (CPP) device. This handheld instrument has a low space requirement (about 21 cm length and 4 cm diameter) and is directly plugged into a standard 230 V-socket. This device facilitates the generation of cold atmospheric plasmas using the piezoelectric effect provided by a lead zirconate titanate ceramic (LZT). The piezoelectric direct discharge achieved by this device offers a voltage transformation ratio exceeding 1000, e.g. amplifying 15 V to 15 kV. Two distinct nozzles can be coupled at the end of the LZT ceramic: either a nearfield-nozzle to generate a dielectric barrier discharge (Pz-DBD) or a needle-nozzle to generate a corona discharge (Pz-CD) [25]. The Pz-DBD source (scheme in **Fig. 1b**) features a floating plane electrode facing the LZT ceramic, establishing a galvanic connection. The addition of a dielectric barrier allows the generation of a DBD inside the nozzle and another DBD occurring between the dielectric and the substrate located a few 2–3 millimeters below. A homemade flow driver (0.4 slm) is inserted between the Pz-DBD and the culture well to confine the helium-air mixture. The Pz-CD source (scheme in **Fig. 1c**) consists of a pin electrode coupled with another homemade flow driver to supply gas mixtures. There, the substrate is located 4–5 mm below the pin. If micro-discharges can be directly generated in ambient air in the Pz-DBD configuration, the Pz-CD requires the injection of helium to generate cold plasma [28]. In order to compare these configurations on the same basis, a very low flow rate of helium (0.5 slm for Pz-CD and 0.4 slm for Pz-DBD) was injected in the gaseous environment confined by the helium flow driver and the well of the culture plate (**Fig. 1b-c**). In the two configurations, mass spectrometry measurements indicate a helium/air ratio of 5.6% along all plasma operation time.

**Operating conditions:** in both cases, the 6-well plate is placed on a laminar flow cabinet during processing, with the metal base connected to the ground. Wells are filled with 1.2 mL of culture medium for Pz-DBD and 1.4 mL for Pz-CD to optimize the treatments. Most of the experiments are carried out for the following CPP exposure conditions: 0, 0.5, 1, 2, 3 and 5 min.

## 4.2. Temperature and evaporation of cell culture medium

**Temperature:** an infrared camera (Jenoptik & Infratec VarioCAM® HD head 680/30) is employed to measure the temperature of the culture medium at the bottom of a grounded 6-well plate, without cell. A macro-lens (Converter Jenoptik ×0.2) focuses on the bottom of the well, where cells are typically placed during plasma treatments. Infrared images are recorded in the medium every 30 s with the following scheme: 30 s without plasma nor helium; 30 s of helium injection but no plasma, 5 min of CPP exposure and 9 min of cooling without plasma nor helium. A radial temperature profile is then extracted at each current treatment time to illustrate its distribution in the medium.

**Evaporation:** the medium is weighted using a Sartorius ENTRIS 124i-1 S scale accurate at 0.1 mg to determine the evaporation inside a well of 6-well plates filled with its current volume of medium for CPP exposure from 0 to 5 min. Mass are measured before CPP treatment, immediately post-treatment and 24 h after incubation at 37 °C.

## 4.3. CPP electric properties and deposited power on a target

A digital oscilloscope (Teledyne Lecroy Company, Wavesurfer 3054) is coupled to a voltage probe (Teledyne LeCroy PP020 10:1) and a current monitor (Pearson Electronics Company, Pearson 2877) to acquire electric signals. Voltage and current are measured on a target mimicking human body impedance designed on previous work conducted by Judée et al. [40]. The target corresponds to a metal plate connected to a derivative circuit constituted by a resistor ($R_T$ = 1.5 k$\Omega$) and a capacitor ($C_T$ = 100 pF) and facing each CPP. While the target voltage is measured between the metal plate and the ground, the target current is measured before the ground (**Fig. 1d**). Signals, sampled at a rate of 2 GS/s, are smoothed by weighted adjacent-averaging with 100 points. Deposited power is calculated from five periods of the integrated instantaneous power, i.e. by integrating the product of target voltage by target current.

## 4.4. Reactive species analysis

**Optical emission spectroscopy:** an optical emission spectrometer (OES – Andor SR-750-B1-R) in Czerny Turner configuration collects the radiative emissions of plasma. Its adjusted parameters are as follows: a focal length of 750 mm, a 1200 grooves/mm grating, blazed at 500 nm and a range from 250 to 800 nm. This device is coupled to an ICCD camera (Andor Istar DH340T) and an optical fiber (Leoni fiber optics SR-OPT-8014) of 100 μm diameter. A converging lens (ThorLabs LA4380-UV, f = 100 mm) is added between plasma and optical fiber to collect a maximum of plasma emission. OES characterization is performed with culture medium below CPP; measurements are acquired orthogonally to plasma axis at a 1 mm-height from the culture medium surface.

**Colorimetric assays:** nitrite ($NO_2^-$) and hydrogen peroxide ($H_2O_2$) concentrations are measured by colorimetric assays respectively using Griess reagent (Sigma Aldrich, Saint-Quentin Fallavier, France) and Titanium Oxysulfate IV ( from Sigma-Aldrich, Saint-Quentin Fallavier, France), to check whether reactive species are produced by plasma in the culture medium, without cell. CPP treatments are performed as usual from 0 to 5 min in a 6-well plate filled with current volumes of culture medium. From each treated sample, 25 µL are transferred in a 96-well plate with 175 µL of culture medium and 50 µL of Griess reagent for nitrite quantification and 200 µL of the sample is transferred in another well with 50 µL of titanium oxysulfate IV for hydrogen peroxide quantification. Their proportion in culture media are recorded by an absorption spectrophotometer (Biotek Cytation 3 instrument), at 548 nm for nitrite species and at 409 nm for hydrogen peroxide.

## 4.5. Cell culture and plasma treatment

**Cell culture:** human intrahepatic biliary duct cell line HuCCT-1 (kindly provided by G. Gores, Mayo Clinic, MN, USA) and extrahepatic biliary duct cell line EGI-1 (Leibniz Institute DSMZ-German Collection of Microorganisms and Cell Cultures GmbH, Braunschweig, Germany) are cultured in DMEM supplemented with 1 g/L glucose, 10 mmol/L HEPES, 10% fetal bovine serum, antibiotics (100 UI/mL penicillin and 100 mg/mL streptomycin) and antimycotic (0.25 mg/mL amphotericin B). Cell lines are routinely screened for the presence of mycoplasma and authenticated for polymorphic markers to prevent cross-contamination.

**Plasma treatment:** cells are plated in 6-well plates at the density of $2 \times 10^5$ cells/well. Twenty-four hours later, the medium is replaced by fresh culture medium, and cells are exposed to CPP. Cell layer distribution and viability are assessed at the current operating conditions: 0, 0.5, 1, 2, 3 and 5 min. The other experiments are conducted at 2 and 3 min for HuCCT-1 cells and 3 and 5 min for EGI-1 cells.

## 4.6. Cell viability

The cell viability is determined by the crystal violet assay, 24 h after treatment and incubation at 37 °C. Absorbance is quantified with a Tecan Microplate Reader at 595 nm.

## 4.7. Staining and immunostaining

Cells from one half of the plates are fixed 15 min immediately after plasma treatment: cells in each well are washed in cold PBS 3 times before being fixed with 4% paraformaldehyde (PFA) at 4 °C. The fixed cells are permeabilized with triton 0.1% at room temperature (RT) for 15 min. The other half is incubated after plasma treatment, for 24 h at 37 °C before the following experiments.

Nuclear staining for cell layer distribution analysis: after 3 PBS washes, 4',6-diamidino-2-phenylindole (DAPI − Thermo Fisher Scientific) is added and cells are incubated for 5–10 min. Another PBS wash is done before storing them in PBS at 4 °C. Observations by fluorescence are performed with a Biotek Cytation™ 3 cell imaging multi-mode reader. Data treatments are realized with a Python program and GIMP2 software to highlight the contrasts of the region of interest and to count the nuclei as pixel aggregates.

**Cytoskeleton analysis:** the filamentous actin (F-actin) are labeled with 1X Phalloidin iFluor 488 conjugate in 1% bovine serum albumin (BSA − Abcam) for 90 min in the dark. The stained cells are washed, nuclei counterstained in a DAPI solution (Thermo Fisher Scientific) for 5–10 min and stored in PBS protected from light. Fluorescence images are acquired with IX83 Evident microscope (× 40) and analyzed with FIJI software.

**Immunofluorescence staining:** the fixed cells are permeabilized in triton 0.2% for 15 min and incubated in a blocking solution containing PBS with 1% BSA and 10% goat serum for 1 h at RT. Cells are then incubated with the γH2AX antibody solution diluted in 5% BSA/0.1% PBS-Tween in a moisture chamber overnight at 4 °C. Twenty-four hours later, after 3 PBS washes for 5 min each, cells are incubated for 1 h at RT with secondary fluorescent antibody diluted in PBS-Tween 0.1% protected from light (**Table 1**). The cells nuclei are counterstained in a DAPI solution (Thermo Fisher Scientific) for 5–10 min. Finally, the cells are rinsed in PBS 3 times for 5 min each before being mounting and cover slipped using a mounting medium. Fluorescence images are acquired with IX83 Evident microscope (× 20) and analyzed with FIJI software.

*Table 1. Primary antibodies used for immunodetection and probe. M, mouse; R, rabbit; WB, western blot; IF, immunofluorescence; CST, Cell Signaling Technology.*

| Name | Spe-cies | Clone | Ref. | Manu-facturer | Dilution |
|---|---|---|---|---|---|
| GAPDH | M | 6C5 | sc-32233 | Santa Cruz | 1:5000 (WB) |
| γH2AX | R | 20E3 | 9718 | CST | 1:1000 (WB), 1:200 (IF) |
| pCHK1 | R | 133D3 | 2348 | CST | 1:1000 (WB) |
| CHK1 | R | 2G1D5 | 2360 | CST | 1:1000 (WB) |
| pp53 | R | / | 9284 | CST | 1:1000 (WB) |
| p53 | M | DO-1 | sc-126 | Santa Cruz | 1:1000 (WB) |
| Cleaved caspase 3 | R | 5A1E | 9664 | CST | 1:1000 (WB) |
| Cleaved PARP | R | D64E10 | 5625 | CST | 1:300 (WB) |
| Phalloidin-iFluor 488 | R | / | ab176753 | Abcam | 1:1000 (IF) |

## 4.8. Western blot

Cells are collected 24 h after CPP exposure and washed with ice-cold PBS solution, then lysed in RIPA buffer (Merk) supplemented with a protease and phosphatase inhibitor cocktail (Thermo Fisher Scientific). After 15 min of incubation in ice, the cells are centrifuged for 15 min at 4 °C at 18,000 g. The protein concentrations in the supernatant are determined using the bicinchoninic acid (BCA assay kit − Thermo Fisher Scientific). Absorbance is quantified with a Tecan Microplate Reader at 570 nm. A protein extract of 20 µg is suspended in Laemmli buffer, boiled, subjected to SDS-PAGE and transferred to nitrocellulose membranes. The blots are blocked with TBS, 0.1% Tween-20 containing 5% BSA and incubated with primary antibodies overnight at 4 °C. The membrane is washed 3 times and then incubated with secondary HRP-linked antibody for 1 h at RT. The signals are revealed using an enhanced chemiluminescence (ECL Prime) kit and the bands are visualized with an Imager iBright. GAPDH is used as an internal reference. The primary and secondary antibodies used are listed in Table 1 and previously validated by the team [19].

## 4.9. RNA and reverse transcription PCR

The total RNA from cells is extracted 24 h after CPP treatment using the Nucleospin RNA kit (MACHEREY-NAGEL). The quantity and quality of RNA are determined spectroscopically using a nanodrop (Thermo Fisher Scientific). Total RNA (1 µg) is utilized to synthesize cDNA using random primers and SuperScript II reverse transcriptase (Thermo Fisher Scientific) according to the manufacturer's protocol. The cDNA samples are utilized for RT-qPCR using SYBR Green Reagent (Roche Diagnostics) and specific primers (**Table 2**), on a LightCycler 96 Roche Instrument. All RT-qPCRs are performed in duplicate. The $2 - \Delta\Delta CT$ method is employed for relative gene expression analysis, with GAPDH, as an internal reference.

*Table 2. Human primers used for quantitative real-time PCR*

| Gene | Protein | | Primers |
|---|---|---|---|
| MSRB3-2 | Methionine sulfoxide reductase B3 | Forward (5' → 3') | CGGTTCAGGTTGGCCTTCATT |
| | | Reverse (5' → 3') | GTGCATCCCATAGGAAAAGTCA |
| SOD1 | Superoxide dismutase 1 | Forward (5' → 3') | GGTGGGCCAAAGGATGAAGAG |
| | | Reverse (5' → 3') | CCACAAGCCAAACGACTTCC |
| SOD2 | Superoxide dismutase 2 | Forward (5' → 3') | GGAAGCCATCAAACGTGACTT |
| | | Reverse (5' → 3') | CCCGTTCCTTATTGAAACCAAGC |
| CAT | Catalase | Forward (5' → 3') | TGGAGCTGGTAACCCAGTAGG |
| | | Reverse (5' → 3') | CCTTTGCCTTGGAGTATTTGGTA |
| HMOX1 | Hemeoxygenase1 | Forward (5' → 3') | GGCAGAGGGTGATAGAAGAGG |
| | | Reverse (5' → 3') | AGCTCCTGCAACTCCTCAAA |
| GAPDH | GAPDH, H | Forward (5' → 3') | AGCCACATCGCTCAGACAC |
| | | Reverse (5' → 3') | GCCCAATACGACCAAATCC |

## 4.10. Membrane permeability

Twenty-four hours after cell plating, the medium is replaced by fresh culture medium containing a green fluorescent YO-PRO™-1, a cell-impermeant dye of 630 Da (already used for studying permeability [41]), at a final concentration of 0.5 µmol/L. Cells are exposed to CPP and protected from light during 15 min. In parallel, negative and positive controls of membrane permeability are prepared by incubating cells with or without triton X-100, one of the most widely used non-ionic surfactants for permeabilizing membranes of living cells. The medium is replaced by a triton solution 10% (Thermo Fisher Scientific) at a final concentration of 0.18 mmol/L for 15 min at 4 °C (positive control) or by PBS (negative control). After 3 PBS washes, the YO-PRO™-1 solution is added at a final concentration of 0.5 µmol/L in a fresh culture medium for 15 min at RT protected from light. Cells are washed in PBS 3 times and fixed with 4% PFA for 15 min at 4 °C protected from light. Fluorescence images are acquired with IX83 Evident microscope (× 10) and analyzed with FIJI software.

## 4.11. Statistics

The results are analyzed using the GraphPad Prism 5.0 statistical software (GraphPad Software, San Diego, CA, USA) and Origin 2019b software (OriginLab, Northampton, USA). Data are shown as means ± standard error of the mean (s.e.m.). For comparisons between two groups, nonparametric Mann–Whitney test are used (*, $p < 0.05$; **, $p < 0.01$; ***, $p < 0.001$).

# 5. Data availability

All datasets generated and analysed during the current study are available from the corresponding authors on reasonable request.

# 7. Acknowledgements

The authors would like to thank Pr. Sophie Thenet, Sorbonne Université, INSERM, Centre de Recherche Saint-Antoine, CRSA; Paris Center for Microbiome Medicine (PaCeMM) FHU, APHP; EPHE, PSL University, Paris, France, for her help in the analysis of cell cytoskeleton . L.F. and T.D. received financial support from ITMO Cancer of Aviesan, INCa and FRM, funds administered by Inserm, CNRS and Sorbonne Université (HRHG-MP22-039, PCSI: 23CP058-00, FRM PMT202306017441). A.P. received a scholarship for 3 years from the French Ministry of Higher Education and Research to complete his doctorate at Sorbonne Université, Paris, France.

# 8. Author contributions

M.S., B.L., T.D and L.F., conceptualization; M.S. B.L., methodology; M.S., B.L., T.D and L.F., validation; M.S., I.H. and B.L., formal analysis; M.S., B.L., I.H., H.D., A.P. and A.C., investigation; T.D, R.M. and L.F., resources; M.S., I.H., B.L. and R.M., data curation; M.S., B.L. and L.F., writing – original draft preparation; M.S., B.L., T.D., L.F., and A.P., writing – review & editing; M.S., B.L., I.H., T.D., and L.F., visualization; T.D. and L.F., supervision; M.S. and B.L., project administration; T.D. and L.F., funding acquisition.

# 9. Competing interests

The authors declare no competing interests.